\documentclass[twocolumn,showpacs,amsfonts,amsmath,amssymb,aps,pra,floatfix]{revtex4}
\usepackage{graphicx}
\usepackage{bm}
\begin{document}
\title{Trojan states of electrons guided by Bessel beams}\thanks{This work combines two topics: Bessel beams of electromagnetic radiation and Trojan states of electrons. To both of these research areas Joe has made decisive contributions \cite{dme,bke}.}
\author{Iwo Bialynicki-Birula}
\email{birula@cft.edu.pl}\affiliation{Center for Theoretical Physics, Polish Academy of Sciences\\Al. Lotnik\'ow 32/46, 02-668 Warsaw, Poland and\\
Institute of Theoretical Physics, Warsaw University}
\author{Zofia Bialynicka-Birula}\affiliation{Institute of Physics, Polish Academy of Sciences,\\ Al. Lotnik\'ow
32/46, 02-668 Warsaw, Poland}
\author{Bartosz Chmura}\affiliation{College of Science, Cardinal Stefan Wyszynski University, Warsaw, Poland}

\begin{abstract} 
Previous work [I. Bialynicki-Birula, Phys. Rev. Lett. {\bf 93}, 20402 (2004)] is extended to cover more realistic examples of electromagnetic waves, viz. the Bessel beams. It is shown that electrons may be guided by a Bessel beam with nonvanishing orbital angular momentum. The mechanism for trapping the electrons near the electromagnetic vortex line of such a wave field is the same as for the Trojan states of Rydberg electrons produced by a circularly polarized wave. The main difference is that in the present case the transverse motion of electrons in a beam is confined under the action of the electromagnetic wave alone, no additional attraction center is required. We also discuss briefly the motion of electrons in Neumann and Hankel beams.
\end{abstract}
\maketitle
\section{Introduction}

The purpose of this paper is to show that Bessel beams of electromagnetic radiation (described in detail in the textbook by Stratton \cite{stratton}) may serve as beam guides for charged particles. The confining mechanism in the transverse direction can be explained as due to an interplay between the Lorentz force and the Coriolis force in the frame rotating with the electromagnetic wave. Exact analytic solutions of the Lorentz, Schr\"odinger, Klein-Gordon, and Dirac equations describing beams of charged particles moving in the presence of an electromagnetic wave with a vortex line have been presented in \cite{ibb} for a special, very simple form of electromagnetic wave carrying angular momentum. These fields are not realistic since the electric and magnetic fields grow without bound with the distance from the vortex line. However, the motion of particles in such Maxwell fields helps to understand the confinement mechanism of particles by electromagnetic vortices. In addition, these simple solutions approximate very well more realistic solutions in the vicinity of vortex lines. In the present paper, we shall show that the same confining mechanism is responsible for guiding electrons inside Bessel beams of electromagnetic field. Bessel beams are still not fully realistic because the field vectors fall off too slowly to make the energy finite, but they are much closer to the physical reality.

Bessel beams of light were produced for the first time by Durnin, Miceli, and Eberly \cite{dme} using an annular slit. Later, Bessel beams were produced also by other methods \cite{vtt,jab,ps,erd,arif,salo,melt,ang}. In order to trap electrons, as we shall explain in the present paper, higher order Bessel light beams are more useful. They were produced first by an axicon \cite{ad} and later in biaxial crystals \cite{king}

\section{Bessel beams of electromagnetic radiation}

Bessel beams appear in a natural way as solutions of Maxwell equations in cylindrical coordinates \cite{stratton}. These solutions are conveniently described using the (differently normalized) Riemann-Silberstein \cite{weber,sil,app,pio} vector $\bm F$,
\begin{eqnarray}\label{rs}
{\bm F} = {\bm E}+ ic{\bm B}.
\end{eqnarray}
With the use of the complex vector $\bm F$, we may rewrite all four Maxwell equations as two equations
\begin{eqnarray}\label{max}
i\partial_t{\bm F} = c\nabla\times{\bm F},\;\;\;\nabla\!\cdot\!{\bm F} = 0.
\end{eqnarray}
The separation of the complex vector ${\bm F}$ into its real (electric) and imaginary (magnetic) parts will be needed when writing down the equations of motion for electrons.

Since we are interested in the beam-like fields, we shall seek the solution of (\ref{max}) in the form
\begin{eqnarray}\label{form}
{\bf F}(x,y,z,t) = e^{\varepsilon i(k_z z -\omega t)}{\tilde{\bf F}(x,y)},
\end{eqnarray}
where $\varepsilon = \pm 1$. Substituting this Ansatz into the Maxwell equations (\ref{max}), we obtain
\begin{eqnarray}\label{max1}
\varepsilon\,\omega\left(\begin{array}{c}
 {\tilde F}_x\\{\tilde F}_y\\{\tilde F}_z\end{array}\right) = c\left(\begin{array}{c}
 \nabla_y {\tilde F}_z - i\varepsilon k_z {\tilde F}_y\\i\varepsilon k_z {\tilde F}_x-\nabla_x {\tilde F}_z\\\nabla_x {\tilde F}_y-\nabla_y {\tilde F}_x\end{array}\right).
\end{eqnarray}
From the first two equations we may determine ${\tilde F}_x$ and ${\tilde F}_y$ in terms of a single complex function $\psi(x,y)$
\begin{subequations}\label{fxyz}
\begin{eqnarray}
{\tilde F}_x &=& \frac{ ck_z\nabla_x\psi(x,y)-i\omega\nabla_y\psi(x,y)}{ck_\perp^2},\\
{\tilde F}_y &=& 
\frac{ck_z\nabla_y\psi(x,y) +i\omega\nabla_x\psi(x,y)}{ck_\perp^2},\\
{\tilde F}_z &=& -i\varepsilon\psi(x,y),
\end{eqnarray}
\end{subequations}
where $k_\perp = \sqrt{\omega^2/c^2 - k_z^2}$. Upon substituting these formulas into the third equation, we obtain the Helmholtz equation in 2D that must be satisfied by $\psi$
\begin{eqnarray}\label{helm}
(\nabla_x^2+\nabla_y^2 + k_\perp^2)\psi(x,y) = 0.
\end{eqnarray}
{\em Every solution} of this equation gives rise to a non-diffracting beam. Various analytic solutions may be obtained by separating the variables.

There are three coordinate systems which allow for the separation of variables: polar, elliptic, and parabolic coordinates (cf., for example, Ref. \cite{moon}). The separation of variables in elliptic and parabolic coordinates in the Helmholtz equation leads to Mathieu and Weber functions, respectively. The corresponding nondiffracting beams look quite intriguing but they seem to be very difficult to produce in reality. In the present paper we shall restrict ourselves to the separation of variables in polar coordinates that leads to Bessel functions. In the degenerate case, when $k_\perp=0$, the Helmholtz equation reduces to the Laplace equation which separates in many other coordinates \cite{moon} and has a plethora of solutions. In particular, every analytic function of either $x+iy$ or $x-iy$ is a solution.

The function $\psi$ for the Bessel beam will be chosen in the form
\begin{eqnarray}\label{psi}
\psi(x,y) = E_0(x+iy)^M\;\frac{J_M(k_\perp\rho)}{\rho^M},
\end{eqnarray}
where $E_0=cB_0$ is the field amplitude measured in units of the electric field.

The Bessel beam may be characterized by four ``quantum numbers'' $k_z, k_\perp, M$, and $\varepsilon$. The meaning of these numbers in terms of the associated eigenvalue problems is discussed in the Appendix. According to Eqs.~(\ref{fxyz}), the solution of Maxwell equations (\ref{max}), characterized by these four numbers, has the form
\begin{widetext}
\begin{eqnarray}\label{cyl1}
{\bm F}_{\{k_z k_\perp\!M\,\varepsilon\}} = E_0\,e^{i\varepsilon(k_z z -\omega t)}\left(\begin{array}{c}
a_+(x+iy)^{M-1}\;\frac{J_{M-1}(k_\perp\rho)}{\rho^{M-1}}
+a_-(x+iy)^{M+1}\;\frac{J_{M+1}(k_\perp\rho)}{\rho^{M+1}}
\\
ia_+(x+iy)^{M-1}\;\frac{J_{M-1}(k_\perp\rho)}{\rho^{M-1}}
-ia_-(x+iy)^{M+1}\;\frac{J_{M+1}(k_\perp\rho)}{\rho^{M+1}}
\\
-i\varepsilon (x+iy)^M\;\frac{J_M(k_\perp\rho)}{\rho^M}
\end{array}\right),
\end{eqnarray}
\end{widetext}
where $\omega=c\sqrt{k_z^2+k_\perp^2}$, $a_\pm = (\omega/c\pm k_z)/2k_\perp$, and $\rho = \sqrt{x^2+y^2}$. In the derivation, we used the formulas $(x\nabla_x + y\nabla_y)=\rho\partial_\rho$, $(x\nabla_y - y\nabla_x)=\partial_\phi$, and also the relations between the Bessel functions and their derivatives $\partial_\rho J_{M}(k_\perp\rho) = \mp MJ_{M}(k_\perp\rho)/\rho \pm k_\perp J_{M\mp 1}(k_\perp\rho)$.

Since Bessel beams carry angular momentum, the electric and magnetic fields rotate as we move around the beam center (the $z$-axis). Moreover, the $z$-axis is at the same time the vortex line (except, when $M=0$) according to the general definition proposed in \cite{bb}.

The Bessel beam for $M=2$ will play a special role in our analysis because it is directly related to our earlier work. Namely, the limit of the Bessel beam with $M=2$, when $k_\perp\to 0$, is the following solution of the Maxwell equations
\begin{eqnarray}\label{msol}
\left(\begin{array}{c}
F_x\\
F_y\\
F_z
\end{array}\right) 
= \frac{E_0 k}{2}\,e^{ik(z - ct)}\left(\begin{array}{c}
(x+iy)\\
i(x+iy)\\
0
\end{array}\right).
\end{eqnarray}
This electromagnetic field is a good (but not uniform) approximation to the Bessel beam for $M=2$ in the region where $k_\perp\rho$ and $k_\perp^2z/k_z$ are much smaller than 1. It is the simplest example of a solution of Maxwell equations characterized in \cite{atop} as ``a vortex line riding atop a null solution'' (null solution means that ${\bf E^2}-c^2{\bf B^2}=0$ and ${\bf E}\!\cdot\!{\bf B}=0$). Electromagnetic wave (\ref{msol}) is not a plane wave but it has the properties found before only for plane waves. As has been shown in \cite{ibb}, one may find analytic solutions of the Lorentz equations of motion of a charged particle in this field and also analytic solutions of the Schr\"odinger, Dirac, and Klein -Gordon equations. In the present work we have used this exactly soluble case as a guide in our study of the particle's motion in a Bessel beam.

\section{Motion of charged particles in a Bessel beam}

We shall analyze the motion of a charged particle in a Bessel beam in a relativistic formulation in view of possible applications to highly energetic electrons. The equations of motion are in this case most conveniently expressed in terms of derivatives (denoted by dots) with respect to the proper time $\tau = \int dt\sqrt{1-{\bf v}^2/c^2}$
\begin{eqnarray}\label{lorentz0}
m\,{\ddot X}^{\mu}(\tau)= e\,f^{\mu\nu}(X(\tau)){\dot X}_{\nu}(\tau).
\end{eqnarray}
The trajectory is described by four functions of $\tau$
\begin{eqnarray}\label{eqrel}
X^{\mu} = \{\xi(\tau),\eta(\tau),\zeta(\tau),\theta(\tau)\} = \{{\bf X}(\tau),\theta(\tau)\}.
\end{eqnarray}
The equations of motion to be solved, in the three-dimensional notation have the form
\begin{subequations}\label{eqnr}
\begin{eqnarray}
{\ddot{\bf X}}(\tau) &=& \frac{e}{m}\!\left(\!\Re{\bf F}(X))\,{\dot{\theta}}(\tau)
+ \frac{\dot{\bf X}(\tau)\times\Im{\bf F}(X)}{c}\!\right)\!,\\
{\ddot{\theta}}(\tau) &=& \frac{e}{m}\,\Re{\bf F}(X)\!\cdot\!\dot{\bf X}(\tau)\,.
\end{eqnarray}
\end{subequations}
In our analysis there will always be a distinguished wave frequency $\omega$ and the corresponding wave-vector length $k=\omega/c$. Therefore, it will be convenient to use $1/\omega$ and $1/k$ as the natural units of time and distance. There are also the characteristic amplitudes of the electric field $E_0$ and of the magnetic field $B_0$. Finally, we shall measure the velocity of electrons in units of $c$. We would like to stress that all values of electron velocities appearing in this paper are the derivatives with respect to the proper time $\tau$. They can exceed the speed of light since they differ from the laboratory velocities by the relativistic factor $\gamma = 1/\sqrt{1-{\bf v}^2/c^2}\geq 1$. In these units, the strength of the interaction of the electron with the electromagnetic field is characterized by a single dimensionless parameter $a$. This dimensionless parameter is known either as the {\em laser-strength parameter} $eE_0/mc\omega$ or the {\em wiggler parameter} $eB_0/m\omega$. Since in our case $E_0=c B_0$, these two numbers as equal. The equations of motion for the dimensionless quantities have the form
\begin{eqnarray}\label{lorentz1}
{\ddot X}^{\mu}(\tau)= a\,{\tilde f}^{\mu\nu}(X(\tau)){\dot X}_{\nu}(\tau),
\end{eqnarray}
where ${\tilde f}^{\mu\nu}=f^{\mu\nu}/E_0$ is the dimensionless field and $\tau$ is measured now in units of $1/\omega$ ($\tau$ is now effectively equal to $\tau\omega$). In the next Section, these equations of motion will be solved numerically for various initial conditions and Bessel beam parameters. All calculations and plots in this work were done with Mathematica \cite{wolfram}.

\section{Electrons guided by Bessel beams}

\begin{figure}
\centering
\includegraphics[width=0.4\textwidth,height=0.9\textheight]{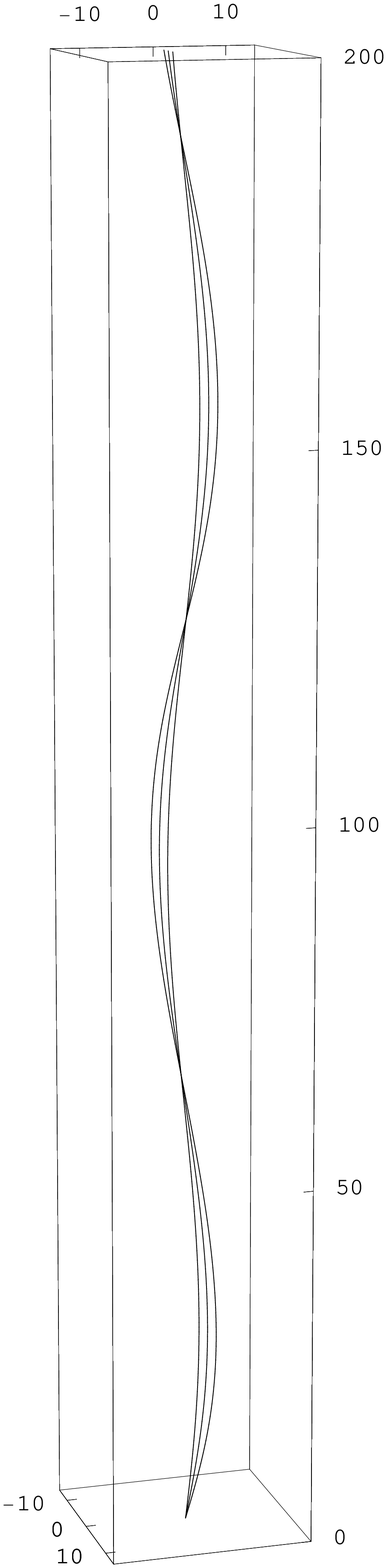}
\caption{The trajectories of electrons guided by the Bessel beam ($M=2$) for $a = 0.0002$, the longitudinal initial velocity $0.002 c$, and three values of the initial transverse velocities: $0.0004\,c$, $0.0008\,c$, and $0.0012\,c$. All distances are measured in units of $c/\omega$.}\label{Fig1}
\end{figure}
\begin{figure}
\centering
\includegraphics[width=0.4\textwidth,height=0.9\textheight]{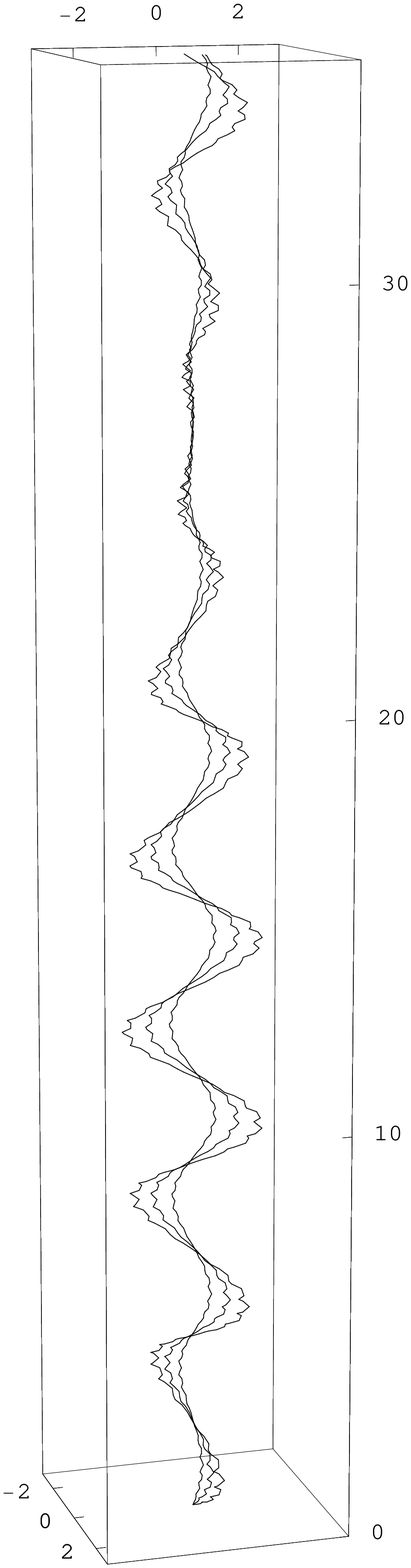}
\caption{The trajectories of electrons guided by the Bessel beam ($M=2$) for $a = 0.14$, the longitudinal initial velocity $0.05\,c$, and three values of the initial transverse velocities: $0.04\,c$, $0.08\,c$, and $0.12 c$. All distances are measured in units of $c/\omega$.}\label{Fig2}
\end{figure}
Bessel beams are capable of trapping and guiding electrons even when they have substantial initial transverse velocities. For example, in the optical case ($2\pi/k=632.8$nm), studied in Ref.~\cite{dme}, electrons with initial transverse velocity as large as 0.0012\,c are trapped by a Bessel beam of moderate intensity of the order of $10^{14}$W/m$^2$. In Fig.~\ref{Fig1} we show the electron trajectories obtained for three different transverse velocities and for $a=0.0002$. In all figures presented in this paper the ratio of the transverse wave vector $k_\perp$ to the longitudinal component of wave vector $k_z$ is 1:100. 

\begin{figure}
\centering
\includegraphics[width=0.45\textwidth]{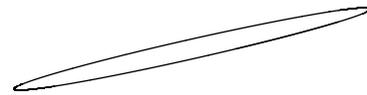}
\caption{The trajectory of a nonrelativistic electron, projected on the $xy$ plane, confined by a Bessel beam. The parameters are the same as in Fig.~\ref{Fig1}.}\label{Fig3}
\end{figure}

\begin{figure}
\centering
\includegraphics[width=0.45\textwidth]{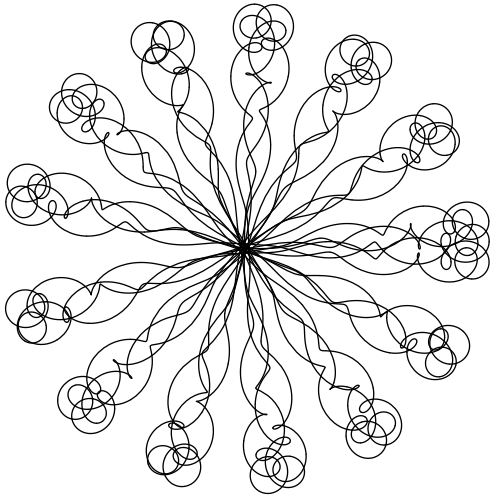}
\caption{The trajectory of a relativistic electron confined by the Bessel beam, projected on the $xy$ plane. The parameters are the same as in Fig.~\ref{Fig2}.}\label{Fig4}
\end{figure}

The trapping of relativistic electrons requires higher values of $a$. This can be achieved either by increasing the intensity or lowering the frequency. In Fig.~\ref{Fig2} we show the trajectories of electrons with initial transverse velocities $0.04\,c$, $0.08\,c$, and $0.12\,c$ and for $a=0.14$. This value of $a$ may, for example, correspond to the microwave frequency $\omega=2\pi\times 10^9$Hz and the intensity $1.5\times 10^9 {\rm W/m}^2$ (calculated from the formula $I~=~ 3.0444\times 10^{-7} a^2\nu^2$, where $I$ is in W/m$^2$ and $\nu$ is in Hz). In Fig.~\ref{Fig3} and Fig.~\ref{Fig4} we show the projection of the electron motion on the $xy$ plane for two different sets of initial conditions. It is clearly seen that the motion in the transverse plane is confined to the vicinity of the vortex line but its details depend very much on the initial data. There is a substantial difference between the slow and fast electrons --- relativistic trajectories exhibit much more elaborate patterns.

\section{Electrons trapped in higher orbits}
\begin{figure}
\centering
\includegraphics[width=0.45\textwidth]{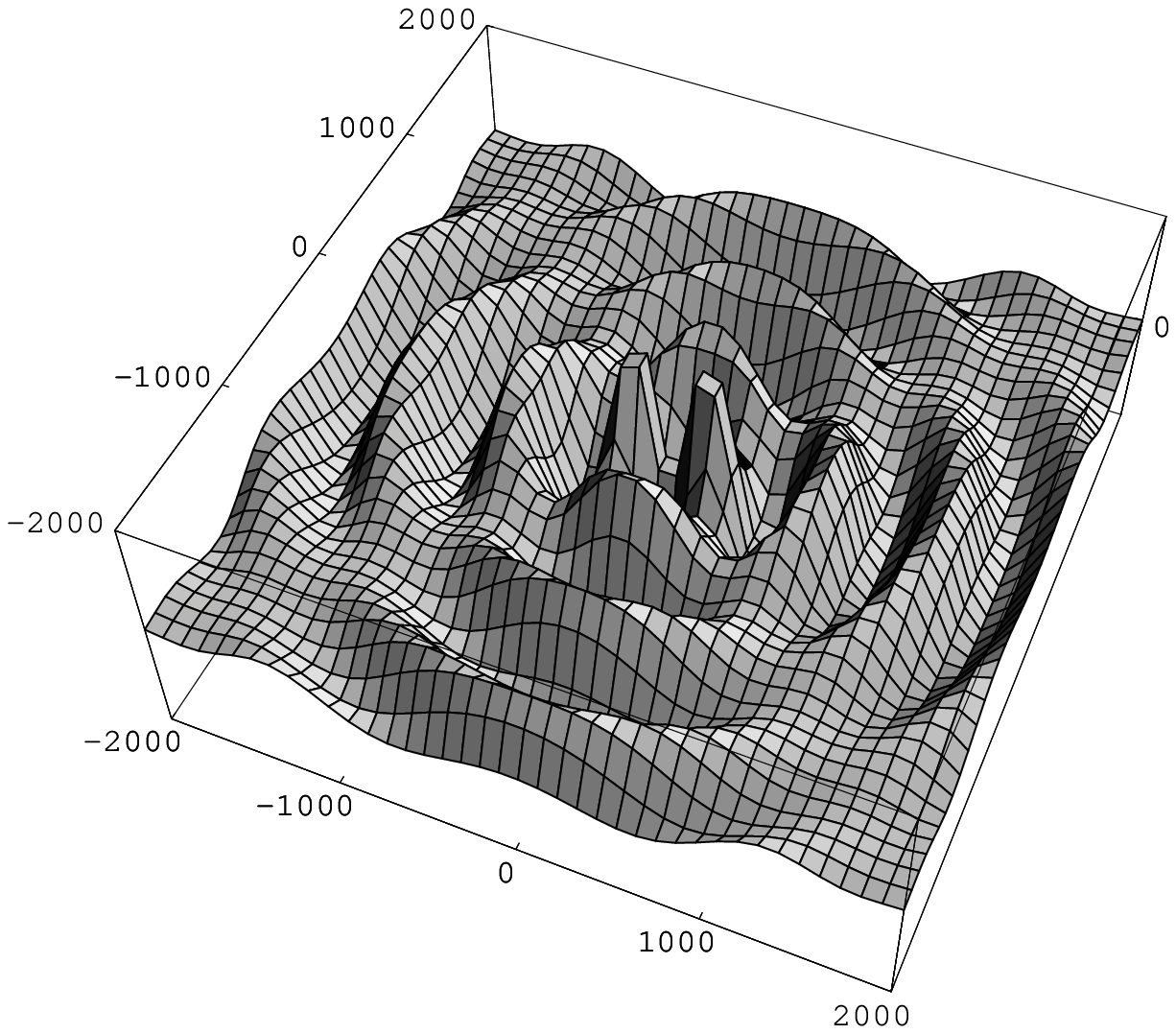}
\caption{The radial component of the electric field $E_\rho$ for the $M=2$ Bessel beam plotted as a function of $x$ and $y$ evaluated at $z=0$ and $t=0$.}\label{Fig5}
\end{figure}
\begin{figure}
\centering
\includegraphics[width=0.45\textwidth]{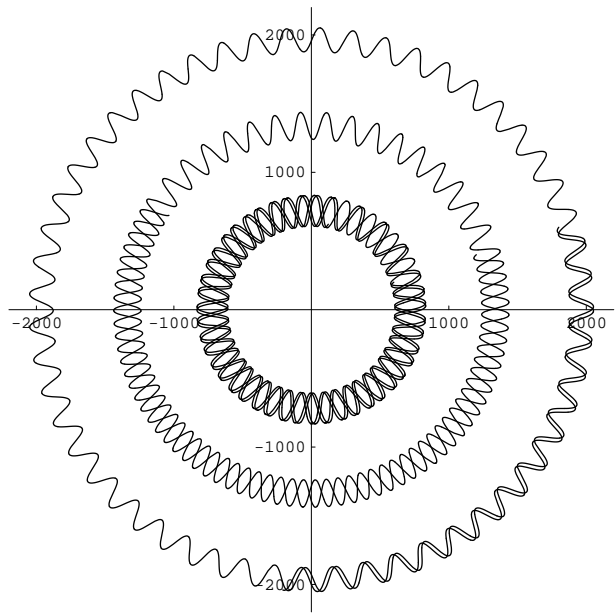}
\caption{These trajectories represent the motion of an electron with the initial velocities $(-0.5,0.5,0)c$, $(-0.1,0.1,0)c$, and $(-0.15,0.15,0)c$ trapped on ``higher orbits'' when $a=0.02$. In order to obtain these orbits, the initial positions of the electron were chosen  sufficiently far from the center as $(1800,600,0)$, $(1200,400,0)$, and $(600,200,0)$, respectively.}\label{Fig6}
\end{figure}

The oscillatory behavior of Bessel functions suggests a possibility of trapping the electrons between two adjacent maxima. Of course, a Bessel beam is far from behaving like a static potential. However, it does produce something like ring-shaped barriers in the transverse direction. To illustrate this point, we show in Fig.~\ref{Fig5} the surface representing the radial component of the electric field. There are regions at distances of about 600, 1200, and 1800 units, where the electric field forms potential wells of a sort where the electrons can perhaps be kept on orbits. The calculation of the electron trajectories in these regions (Fig.~\ref{Fig6}) fully confirms this expectation. We find there stationary (though wiggly) orbits. In contrast to ordinary bound states in static potentials, kinetic energies of electrons trapped in Bessel beams are lower for higher orbits.

\section{Trapping of electrons in Neumann beams}
\begin{figure}
\centering
\includegraphics[width=0.45\textwidth]{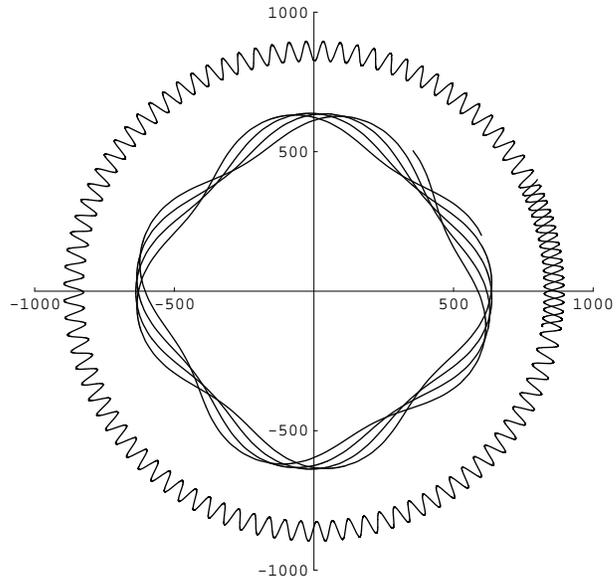}
\caption{These trajectories represent the motion of an electron with the initial velocities $(-0.1,0.1,0)c$ and $(-0.15,0.15,0)c$ trapped on ``higher orbits'' in the $M=0$ Neumann beam when $a=0.02$. In order to obtain these orbits, the initial positions of the electron were chosen  sufficiently far from the center as $(800,400,0)$ and $(600,200,0)$, respectively.}\label{Fig7}
\end{figure}

In addition to regular solutions, the Helmholtz equation has also solutions with singularities. These singular solutions must be excluded if we allow the field to occupy the whole space. However, when portions of space where the singularities occur, due to the presence of some obstacles, are inaccessible, then these singular solutions must, in general, be included to satisfy the boundary conditions. This takes place, for example, in the case of cylindrical coaxial lines. In order to satisfy the boundary conditions we have to include in the solution, in addition to Bessel functions $J_M$ also the Neumann functions $Y_M$ (Bessel functions of the second kind). Since Neumann functions satisfy the same differential equation as the Bessel functions, the solutions of the Maxwell equations describing Neumann beams can be obtained directly from our formulas (\ref{cyl1}) by replacing all Bessel functions by the corresponding Neumann functions. This will give a solution of the Maxwell equations everywhere, except on the line $\rho=0$. Assuming that the vicinity of this line is in some way shielded, we can study the motion of electrons in the region where the electromagnetic field is regular. In Fig.~\ref{Fig7} we show two trajectories of electrons that were obtained under similar conditions as those in Fig.~\ref{Fig6} but for a Neumann beam with $M=0$. We can clearly see that the same mechanism of stabilization in the transverse plane is in place also for Neumann beams.
\begin{figure}
\centering
\includegraphics[width=0.45\textwidth]{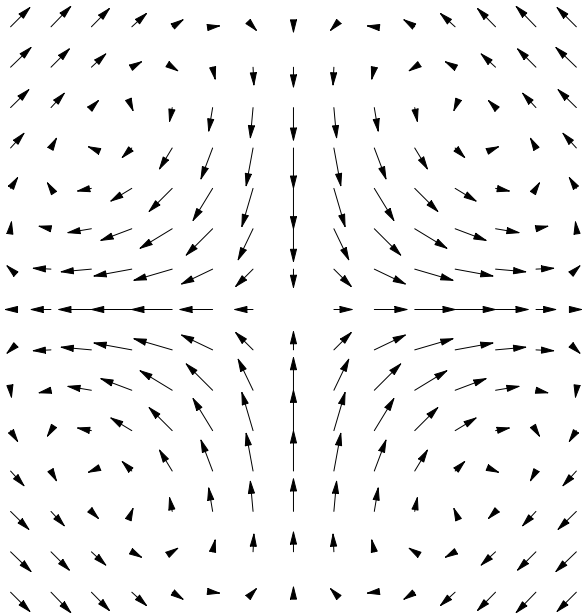}
\caption{The electric field in the Bessel beam for $M=2$, projected on the $xy$ plane and near the vortex line. This field has been evaluated at $z=0$, but for all values of $z$ it has the same general structure.}\label{Fig8}
\end{figure}
\begin{figure}
\centering
\includegraphics[width=0.45\textwidth]{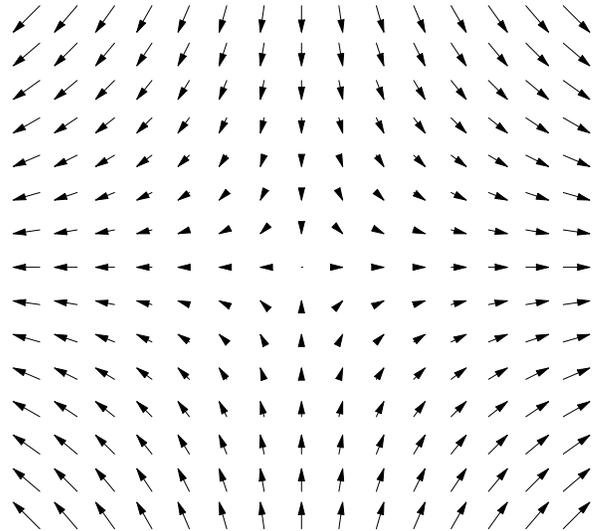}
\caption{The gradient of the saddle surface $z(x,y)=x^2-y^2$, projected on the $xy$ plane. This plot is to be compared with the plot of the electric field in the Bessel beam Fig.~\ref{Fig8}.}\label{Fig9}
\end{figure}
\section{Trojan mechanism of electron trapping}

\begin{figure}
\centering
\includegraphics[width=0.45\textwidth]{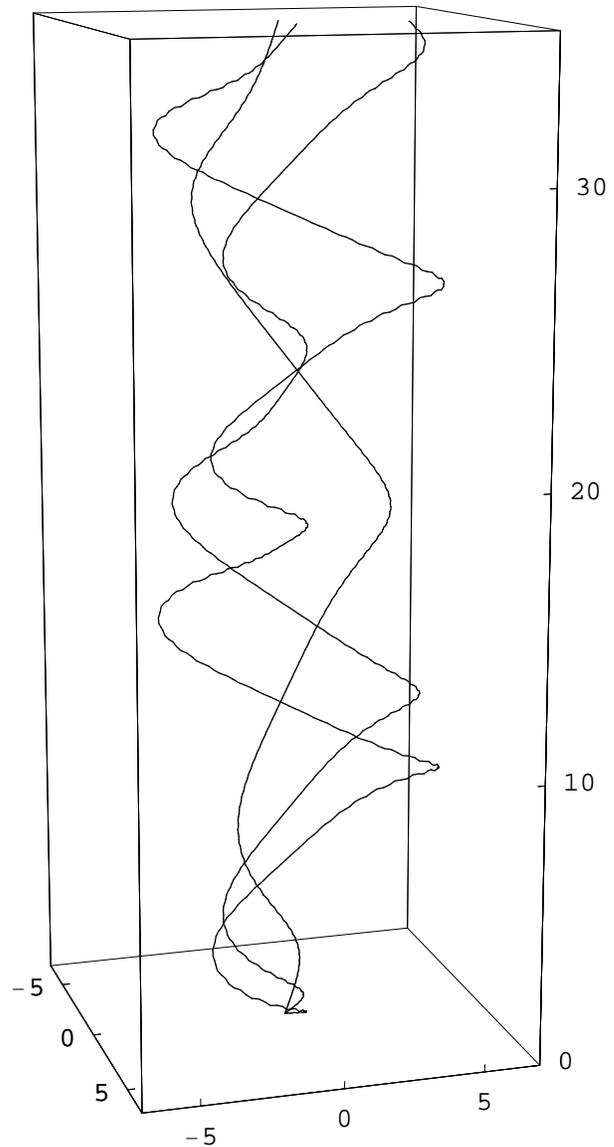}
\caption{The same choice of parameters as in Fig.~\ref{Fig2} but with the electric field of the wave turned off. Note that the amplitude of transverse oscillations is much larger than in Fig.~\ref{Fig11}.}\label{Fig10}
\end{figure}

\begin{figure}
\centering
\includegraphics[width=0.45\textwidth,height=0.9\textheight]{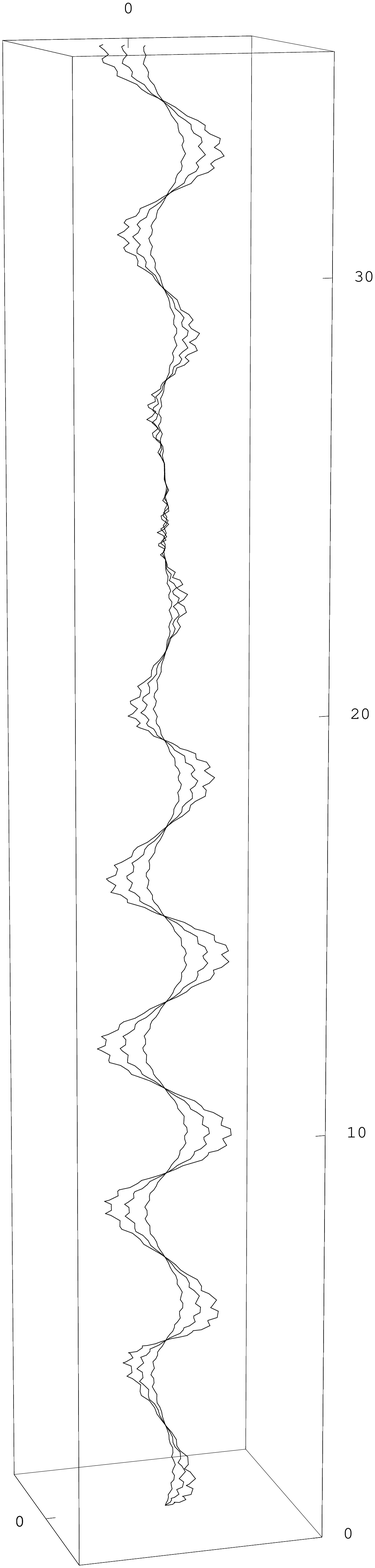}
\caption{The same choice of parameters as in Fig.~\ref{Fig2} but with the magnetic field of the wave turned off. The differences between these trajectories and the ones obtained with the magnetic field on (Fig.~\ref{Fig2}) is barely visible.}\label{Fig11}
\end{figure}

\begin{figure}
\centering
\includegraphics[width=0.45\textwidth]{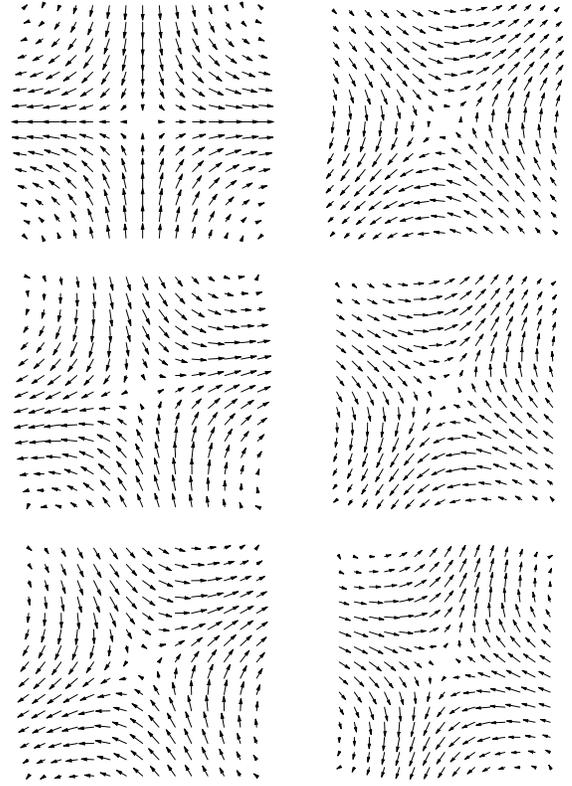}
\caption{The electric field of the Bessel beam for $M=2$, projected on the $xy$ plane is shown as a function of time. This field has been evaluated at $z=0$ for $t$ changing from 0 to $5\pi/12$ in units of $1/\omega$. Time is increasing when we move down in each column. The electric field is rotating clockwise at each point.}\label{Fig12}
\end{figure}

In order to understand the mechanism of electron trapping near the electromagnetic vortex line, we shall rewrite the formulas for the Bessel beams in terms of radial $F_\rho=(xF_x+yF_y)/\rho$ and azimuthal $F_\phi=(-yF_x+xF_y)/\rho$ components of the vector ${\bf F}$ in cylindrical coordinates
\begin{eqnarray}\label{polar}
\left(\!\!\begin{array}{c}
F_\rho\\F_\phi\\F_z\end{array}\!\!\right)\!=\! E_0e^{i\varphi}\!\!\left(\!\begin{array}{c}
a_+J_{M-1}(k_\perp\rho)
+a_-J_{M+1}(k_\perp\rho)
\\
ia_+J_{M-1}(k_\perp\rho)
-ia_-J_{M+1}(k_\perp\rho)
\\
-i\varepsilon J_{M}(k_\perp\rho)
\end{array}\!\!\right)\!,
\end{eqnarray}
where $\varphi=\varepsilon(k_z z -\omega t) + M\phi$. This representation exhibits clearly a screw symmetry of the Bessel beam; changing simultaneously $z$ and $\phi$ in the right proportions $z\to z + \varepsilon M\phi_0/k_z$ and $\phi \to \phi - \phi_0$, leaves the field ${\bf F}$ unchanged. It can also be shown that, as time goes by, at each point in space the tips of the electric and magnetic field vectors follow each other tracing the same ellipse with the frequency $\omega$. The parameters of these ellipses depend on $\rho$ and not on $z$ and $\phi$. Each ellipse lies in a plane determined by its normal vector ${\bf N}$
\begin{eqnarray}\label{normal}
\left(\!\begin{array}{c}
N_\rho\\N_\phi\\N_z\end{array}\!\right) = \left(\!\begin{array}{c}0\\\varepsilon J_{M}(k_\perp\rho)\\a_+J_{M-1}(k_\perp\rho) - a_-J_{M+1}(k_\perp\rho)
\end{array}\!\right).
\end{eqnarray}
We can freeze the motion of the electric and magnetic field vectors by going to a new coordinate frame rotating with the frequency $\omega/M$. The relevant coordinate transformation in the cylindrical coordinate system has the form $\phi\to \phi + \varepsilon\omega t/M$. This transformation eliminates the time variable --- in the rotating frame the electromagnetic forces become time independent. A typical configuration of the electric field is shown in Fig.~\ref{Fig8}. This configuration resembles the forces acting on a ball moving on the saddle surface, shown in Fig.~\ref{Fig9}. Obviously, such a field configuration does not have a stable equilibrium point. However, in a rotating frame, in addition to the Lorentz force, there appear also the Coriolis force and the centrifugal force. These two inertial forces, together with the electric field of the wave, are responsible for the electron trapping in the transverse direction. The magnetic field plays a less important role in the trapping mechanism, as illustrated in Figs.~\ref{Fig10} and~\ref{Fig11}. The present case belongs to the same category of phenomena as the Trojan asteroids \cite{lag,moulton}, the Trojan states of electrons in atoms \cite{bke,bke1} or in molecules \cite{bb1}, and the Paul trap \cite{paul}. In all these systems periodically changing forces lead to a dynamical equilibrium. In the case of Trojan states, the periodical changes of the forces are due to rotation. In the rotating frame the Coriolis force and the centrifugal force create a dynamical equilibrium in an otherwise unstable system. This mechanism is very well illustrated with the use of a mechanical model, a rotating saddle surface, displayed by W. Paul during his Nobel lecture \cite{paul}. In our case, the rotating electric field plays the crucial role. The rotating pattern of the electric field is seen in Fig.~\ref{Fig12}.

\begin{figure}
\centering
\includegraphics[width=0.45\textwidth]{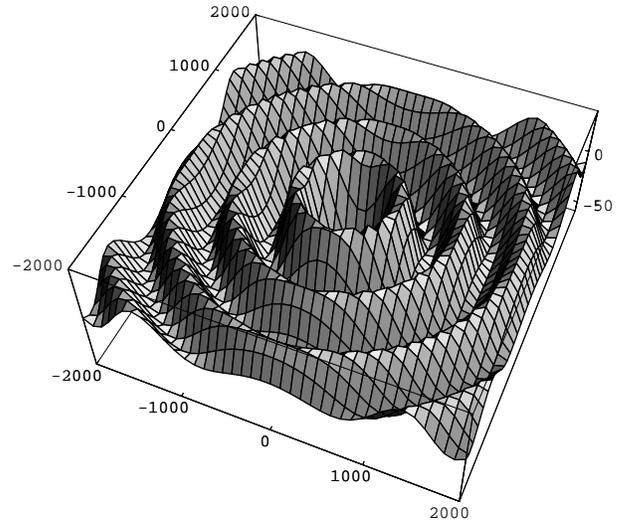}
\caption{The radial component of the electric field $E_\rho$ for the $M=0$ Hankel beam plotted as a function of $x$ and $y$ evaluated at $z=0$ and $t=0$.}\label{Fig13}
\end{figure}

\begin{figure}
\centering
\includegraphics[width=0.45\textwidth]{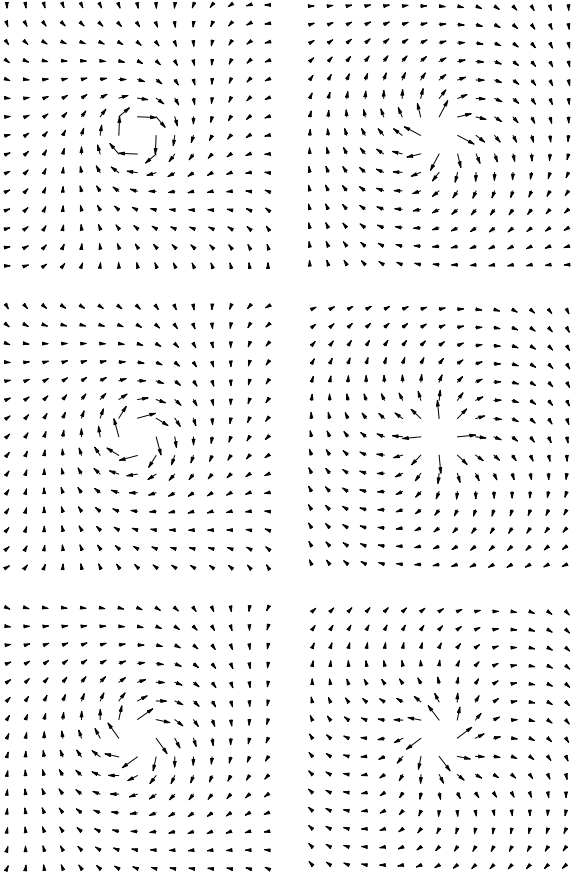}
\caption{The electric field in the Hankel beam for $M=0$, projected on the $xy$ plane is shown as a function of time. This field has been evaluated at $z=0$ for $t$ changing from 0 to $\pi/2$ in units of the wave period. Time is increasing when we move down in each column. The electric field exhibits a totally different pattern than the one in the Bessel beam.}\label{Fig14}
\end{figure}

The same trapping mechanism operates in the case of Neumann beams. However, not all members of the family of Bessel functions can be used for trapping electrons. Special combinations of Bessel and Neumann functions --- the Hankel functions $H_M^{(1)} = J_M +iY_M$ and $H_M^{(2)} = J_M -iY_M$ --- are of interest because they describe outgoing and incoming waves \cite{stratton}. The Hankel beam is described by the Eq.~(\ref{cyl1}) in which all functions $J_M$ are replaced by either $H_M^{(1)}$ or $H_M^{(2)}$. The Hankel beams do not seem to trap charged particles --- we have not been able to find trapped trajectories. This is presumably due to a different structure of the field vectors. In Fig.~\ref{Fig13} we display the radial component of the electric field for the $M=0$ Hankel beam. It clearly has a different character that in the case of a Bessel beam (cf. Fig.~\ref{Fig5}). The ring-shaped barriers are now even more pronounced than those found for a Bessel beam shown in Fig.~\ref{Fig5}. The lack of trapping, however, can be explained by a completely different pattern of the electric field shown in Fig.~\ref{Fig14}. The lines of force now spiral in, instead of forming the saddle pattern of Figs.~\ref{Fig8} and~\ref{Fig9}. 


\section{Scattering of electrons off Bessel beams}

\begin{figure}
\centering
\includegraphics[width=0.45\textwidth]{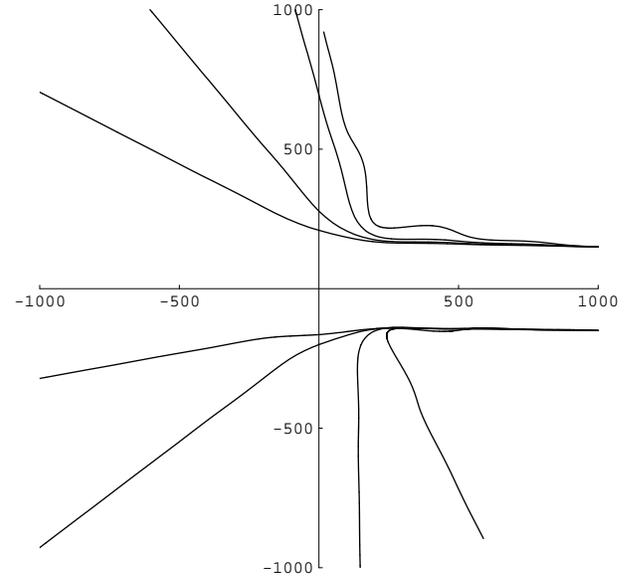}
\caption{Scattering of electrons off a Bessel beam. The trajectories represent scattering of electrons in the plane perpendicular to the beam (the beam center is at the origin) with two initial positions and with the initial velocities $(-0.02,0,0)\,c$, $(-0.03,0,0)\,c$, $(-0.04,0,0)\,c$, and $(-0.05,0,0)\,c$. The strength of the Bessel beam is determined by $a=0.005$.}\label{Fig15}
\end{figure}

Bessel beams are the strongest at the first maximum and the value of the field at subsequent maxima decreases as $1/\sqrt{\rho}$ when we move away. Therefore, we may observe analogs of scattering phenomena by sending electrons from a distance towards the center of the beam. The trajectories of scattered electrons are, however, quite different from those of potential scattering. Some of them curve in an unexpected manner and there is an obvious left-right asymmetry that is due to the rotation of the field around the vortex line (cf. Fig.~\ref{Fig15}).

\section{Acknowledgements}

This research has been partly supported by the KBN Grant 1 P03B 041 26.
\clearpage
\appendix
\section{Labeling mode functions with quantum numbers}

The mode functions of the electromagnetic field are the analogues of the eigenfunctions of a set of operators in quantum mechanics. This analogy becomes even more succinct when the Riemann-Silberstein vector $\bf F$ is treated as a wave function of the photon \cite{pwf}. The four commuting operators whose eigenvalues serve as quantum numbers that characterize the mode functions are in our case: the $z$-component of the wave vector operator (or momentum divided by the Planck's constant) $\hat k_z = -i\nabla_z$, the length squared of the transverse wave vector $\hat k_\perp^2 = -\Delta_\perp = -\nabla_x^2-\nabla_y^2$, the projection on the $z$-axis of the (dimensionless) total angular momentum operator $\hat J_z = \hat L_z + \hat S_z$, and finally the helicity. The total angular momentum vector is a sum of the orbital part ${\hat{\bf L}} = -i{\bf r}\times\nabla$ and the spin part ${\hat{\bf S}}$
\begin{eqnarray}\label{ls}
\hat S_x = \left(\begin{array}{ccc}
0 & 0 & 0\\
0 & 0 & -i\\
0 & i & 0
\end{array}\right),\;\;\hat S_y = \left(\begin{array}{ccc}
0 & 0 & i\\
0 & 0 & 0\\
-i & 0 & 0
\end{array}\right),\nonumber\\\hat S_z = \left(\begin{array}{ccc}
0 & -i & 0\\
i & 0 & 0\\
0 & 0 & 0
\end{array}\right).
\end{eqnarray}
For a plane wave the helicity is associated with the sense of circular polarization (right or left). More generally, the helicity $\varepsilon$ can be defined as the sign of the projection of the angular momentum ${\hat{\bf J}}$ on the direction of the wave vector
\begin{eqnarray}\label{hel}
\hat\varepsilon = {\rm sign}({\hat{\bf J}}\!\cdot\!{\hat{\bf k}})
= {\rm sign}({\hat{\bf S}}\!\cdot\!{\hat{\bf k}}).
\end{eqnarray}
Since the operator ${\hat{\bf S}}\!\cdot\!{\hat{\bf k}}$ is nothing else but the curl
\begin{eqnarray}\label{hel1}
({\hat{\bf S}}\!\cdot\!{\hat{\bf k}})
= \left(\begin{array}{ccc}
0 & -\nabla_z & \nabla_y\\
\nabla_z & 0 & -\nabla_x\\
-\nabla_y & \nabla_x & 0
\end{array}\right),
\end{eqnarray}
we may write the Maxwell equations in the form
\begin{eqnarray}\label{hel2}
i\partial_t{\bm F} = c ({\hat{\bf S}}\!\cdot\!{\hat{\bf k}}){\bm F}.
\end{eqnarray}
It follows from this formula that for monochromatic waves $\varepsilon$ coincides with the sign of the frequency. A Bessel beam ${\bm F}_{\{k_z k_\perp\!M\,\varepsilon\}}$ may, therefore, be determined from the the following set of eigenvalue equations
\begin{eqnarray}\label{eigen}
-i\nabla_z{\bm F}_{\{k_z k_\perp\!M\,\varepsilon\}} = k_z{\bm F}_{\{k_z k_\perp\!M\,\varepsilon\}},\;\nonumber\\
-(\nabla_x^2+\nabla_y^2){\bm F}_{\{k_z k_\perp\!M\,\varepsilon\}} = k_\perp^2{\bm F}_{\{k_z k_\perp\!M\,\varepsilon\}},\nonumber\\
\left(-i(x\nabla_y - y\nabla_x) + {\hat S_z}\right){\bm F}_{\{k_z k_\perp\!M\,\varepsilon\}} = M{\bm F}_{\{k_z k_\perp\!M\,\varepsilon\}},\nonumber\\
{\rm sign}({\hat{\bf J}}\!\cdot\!{\hat{\bf k}}){\bm F}_{\{k_z k_\perp\!M\,\varepsilon\}} = \varepsilon{\bm F}_{\{k_z k_\perp\!M\,\varepsilon\}}.\;\;\;
\end{eqnarray}

\end{document}